\documentstyle[11pt,aaspp4]{article}

\lefthead{}
\righthead{}

\def\einstein{{\it Einstein}}
\def\rosat{{\it ROSAT}}

\def\chandra{{\it Chandra}}
\def\CXO{{\it Chandra X-Ray Observatory}}
\def\eohone{E0102.2$-$7219}
\def\kms{km s$^{-1}$}

\def\myarcsec{\mskip1mu^{\prime\prime}\mskip-7mu.\mskip2mu}
\def\lsim{\hbox{\raise.35ex\rlap{$<$}\lower.6ex\hbox{$\sim$}\ }}
\def\gsim{\hbox{\raise.35ex\rlap{$>$}\lower.6ex\hbox{$\sim$}\ }}

\begin{document}

\title{Electron Heating and Cosmic Rays at a Supernova 
Shock from \chandra\ X-ray Observations of \eohone}

\author{
John P.~Hughes\altaffilmark{1,2},
Cara E.~Rakowski\altaffilmark{1,2}, and
Anne Decourchelle\altaffilmark{2}
}
%\affil{
%Service d'Astrophysique, L'Orme des Merisiers, 
%CEA-Saclay, 91191 Gif-sur-Yvette Cedex France
%}

\altaffiltext{1}{Department of Physics and Astronomy, Rutgers The State
 University of New Jersey, 136 Frelinghuysen Road, Piscataway NJ 08854-8019;
 E-mail: jph@physics.rutgers.edu and rakowski@physics.rutgers.edu
}

\altaffiltext{2}{
Service d'Astrophysique, L'Orme des Merisiers, 
CEA-Saclay, 91191 Gif-sur-Yvette Cedex France
}

\begin{abstract}

In this {\it Letter} we use the unprecedented spatial resolution of
the \CXO\ to carry out, for the first time, a measurement of the
post-shock electron temperature and proper motion of a young SNR,
specifically to address questions about the post-shock partition of
energy among electrons, ions, and cosmic rays.  The expansion rate,
$0.100\% \pm 0.025\% \rm\, yr^{-1}$, and inferred age, $\sim$1000 yr,
of \eohone, from a comparison of X-ray observations spanning 20 years,
are fully consistent with previous estimates based on studies of high
velocity oxygen-rich optical filaments in the remnant.  With a radius
of 6.4 pc for the blast wave estimated from the \chandra\ image, our
expansion rate implies a blast wave velocity of $\sim$6000 km s$^{-1}$
and a range of electron temperatures 2.5 - 45 keV, dependent on the
degree of collisionless electron heating.  Analysis of the
\chandra\ ACIS spectrum of the immediate post-shock region reveals a
thermal plasma with abundances and column density typical of the Small
Magellanic Cloud and an electron temperature of 0.4--1 keV.
The measured electron temperature is significantly lower than the
plausible range above, which can only be reconciled if we assume
that a significant fraction of the shock energy, rather than
contributing to the heating of the post-shock electrons and ions, has
gone into generating cosmic rays.

\end{abstract}


\section{INTRODUCTION}

\par

Supernova remnants (SNRs) are among that rare class of astronomical
objects whose dynamical evolution can be probed directly through both
radial velocity and proper motion measurements.  Together such studies
provide important constraints on the evolution of young SNRs and the
structure of the ejecta and ambient medium.  Furthermore, when its
distance is known, measurement of a remnant's proper motion is
equivalent to a measurement of the velocity of the supernova blast
wave. This fundamental quantity determines the total energy available
to the remnant for partition into bulk kinetic energy of the ambient
medium, shock-heated electrons and ions, and relativistic particles,
i.e., cosmic rays (see Chevalier 1983; Decourchelle, Ellison, \&
Ballet 2000).  Conservation laws provide some constraints on how the
shock energy is divided up into the several forms, but some issues,
such as the fraction of shock energy that goes into cosmic rays 
and the extent of collisionless heating of the electrons, are still
theoretically uncertain and need to be determined observationally.

The brightest X-ray supernova remnant (SNR) in the Small Magellanic
Cloud (SMC), \eohone, is dominated by line emission from highly
ionized atoms of O, Ne, and Mg (Hayashi et al.~1994; Gaetz et
al.~2000) and its X-ray morphology is that of a strongly
limb-brightened shell, roughly 40$^{\prime\prime}$ in diameter (Hughes
1988).  Optical filaments in the remnant are also rich in O and Ne,
extend over a diameter of 24$^{\prime\prime}$ (Dopita et al.~1981),
and display radial velocities that span a range of 6500 \kms\ (Tuohy
\& Dopita 1983).  At the distance to the SMC (60 kpc, e.g., van den
Bergh 2000), the proper motion of the optical filaments should be
$0\myarcsec011$ yr$^{-1}$ or an expansion rate of 0.092\% yr$^{-1}$.
Herein we determine the X-ray expansion rate, equivalent to the
velocity of the shock, to obtain an estimate of the total shock
energy. We then compare this to the electron temperature in the
immediate post-shock region, through fits to spectra isolated by
\chandra's superb spatial resolution, to determine the fraction of
energy in shock-heated electrons.

\section{EXPANSION RATE}

Prior to the launch of \chandra\ the highest spatial resolution images
of \eohone\ were those taken by the \einstein\ and \rosat\ High
Resolution Imagers, hereafter EHRI and RHRI. During Orbital Activation
and Check-out of \chandra, \eohone\ was observed by the back-side
illuminated chip (S3) of the ACIS-S instrument (Garmire 1997). A
processed event file (id no.~1231 created on 1999-09-07T18:39:11) was
obtained from the \chandra\ X-ray Center and analysis was carried out
using standard astronomical software. For spectral analysis (\S 3) the
events were gain-corrected using the appropriate gain map for a focal
plane temperature of $-$100 C.  Events were time filtered to remove
intervals of high background as well as frames with bad or no
aspect. Dead and flickering pixels were also removed.  An ephemeris of
all three observations is given in Table 1.

The EHRI and RHRI each made broad-band X-ray images, but in slightly
different spectral bands. The possibility of spectral variations with
position has been the main factor limiting the widespread use of EHRI
and RHRI images for carrying out precise proper motion studies of
X-ray SNRs.  Now, however, the \chandra\ ACIS-S data provide a
spectral dimension that can be used to take full advantage of the long
time baseline for proper motion studies set by the earlier
observations.

In our approach we slice the ACIS-S data into a large number (50--100)
of images according to observed photon energy using narrow X-ray
spectral bands.  These separate images are then multiplied by the
ratio of effective area between the EHRI (for example) and the
\chandra\ ACIS-S at the photon energy appropriate to the spectral band
of each individual image slice.  The final ACIS-S image, corrected to
the EHRI band in this case, is then the sum of the image slices.  This
is done for both of the HRIs.  Clearly the accuracy of this procedure
depends on how well the various effective area functions are known, an
effect we explore below.

The on-orbit 50\% encircled energy radius for the \chandra\ ACIS-S,
$R_{50\%}=0\myarcsec4$ (Dewey 1999), is about an order of magnitude
smaller than that of the RHRI or EHRI, both of which had $R_{50\%}
\approx 4^{\prime\prime}$ (Giacconi et al.~1979; David et al.~1998).
We can therefore ignore the small \chandra\ point-response-function
(PRF) and convolve the band-corrected ACIS-S image with the
appropriate EHRI or RHRI PRF models. In our preliminary investigations
we found that the parameters of the analytical PRF models (EHRI: Henry
\& Henriksen 1986; RHRI David et al.~1998) needed to be modified
slightly from their published values in order to obtain good fits to
the \eohone\ data. For both datasets it was the largest spatial scale
component of the PRF model that needed adjustment. For the RHRI the
best-fit normalization of the $\sim$30$^{\prime\prime}$-scale
exponential component was 33\% higher than nominal and for the EHRI
the normalization of the $\sim$13$^{\prime\prime}$-scale power-law
component was less than nominal by 22\%. \eohone\ is small enough that
spatial variations in the background, vignetting, or PRF can all be
safely neglected.

The final piece of information required for the expansion study is the
plate scales of the three instruments.  The plate scales of the EHRI
and RHRI are $0\myarcsec4965 \pm 0\myarcsec0012$ per pixel and
$0\myarcsec499 \pm 0\myarcsec001$ per pixel, respectively (David et
al.~1998). For the ACIS-S the plate scale near the center of the field
of view is $0\myarcsec49115 \pm 0\myarcsec00010$ per pixel based on
\chandra\ in-flight observations of the star cluster NGC 2516
(M.~Markevitch 1999, private communication).  This value is within
0.2\% of the pre-flight estimate of $0\myarcsec4920$ per pixel.

The approach we follow for determining the expansion rate is that
described in Hughes (1999).  In brief each HRI image (the ``data'') is
compared to the appropriate band-corrected ACIS-S image (the
``model'') in a fit for parameters using a maximum-likelihood
estimator as the figure-of-merit function.  The fitted parameters are
the relative intensity scale between the two images, the relative
pixel position, the fractional amount of expansion or contraction of
the spatial scale, and the normalization of the largest-scale PRF
component.  The difference in background levels (e.g., ACIS-S vs.\
EHRI or HRI) is fixed at the value determined from source-free
portions of the image and is included as a spatially uniform term. In
the fitting software the expansion or contraction of the ``model''
image is done first using a rebinning scheme that conserves flux, then
the scaled image is convolved with the appropriate instrumental PRF,
and finally the background term is included.  The only interesting
parameter that we obtain from this comparison is the expansion (or
contraction) factor, which tells us the global mean expansion rate of
the remnant.  This procedure explicitly assumes that the fractional
expansion rate is uniform over the entire remnant, both radially and
azimuthally.

Figure 1 shows the fitted results.  \eohone\ has expanded by
$0.74\pm0.31$\% based on comparing the RHRI and ACIS-S data and it has
expanded by $1.98\pm0.64$\% based on comparing the EHRI and ACIS-S
data. The quoted errors are purely statistical at 1 sigma and include
both Poisson error (dominated by the EHRI and RHRI data) as well as
the uncertainty in the plate scales. The two points are described best
by a uniform expansion rate of $0.100\pm0.025$\% yr$^{-1}$. The null
hypothesis that the remnant has remained constant in size can be
rejected at the $4\sigma$ confidence level.

It is extremely unlikely that systematic effects could account for
these results.  For example, the \chandra\ plate scale would have to
be incorrect by more than 1\%, which is 50 times the statistical
uncertainty on the calibration measurement.  We investigated
systematic errors due to uncertainty in the analytical PRF models and
in the effective area functions and the boxes plotted in fig.~1 show
an estimate of their likely effect on the expansion measurements.
Neither is expected to be a dominant source of error.

\section {POST-SHOCK ELECTRON TEMPERATURE}

The eastern limb of \eohone\ in the \chandra\ image (Blair et
al.~2000, Gaetz et al.~2000) displays a smooth, nearly circular, faint
rim of X-ray emission.  Beyond the edge of the rim the surface
brightness drops by more than two orders of magnitude to the average
background level over a distance of order 1$^{\prime\prime}$. Herein
we identify this jump as the SN shock wave plowing into the
ambient medium (hereafter ``blast wave'').

We extracted the spectrum of the blast wave region from within a
partial annulus of thickness $1\myarcsec5$ and outer radius
21$^{\prime\prime}$ that covered the angular range from $-$30$^\circ$
to 155$^\circ$ (angles measured counterclockwise from north).  We
ignored the background since it was negligibly small (less than 1\%).
The spectral data were contained within a single readout node of chip
S3, nevertheless the spectrum included data from several detector
regions with potentially different intrinsic responses to incident
X-rays.  Individual response functions (both effective area functions
and spectral redistribution matrices) were generated for the several
regions covered by the extraction region and then combined, weighting
by the number of X-ray events, into a single global response function.
The results we obtain are insensitive to this procedure:
our fitted parameters are nearly unchanged if we use the response
functions from any individual extraction region.  Because
of the limited statistical precision of our spectrum (which contained
only $\sim$1500 detected events) the uncertainty on derived parameters
is dominated by Poisson counting errors rather than calibration
uncertainties.

Figure 2 plots the blast wave spectrum and best-fit nonequilibrium
ionization model (NEI) (Hughes \& Singh 1994).  The spectrum shows a
well-resolved \ion{O}{8} Ly$\alpha$ line at 0.653 keV, a blend of
\ion{Ne}{9} and \ion{Ne}{10} K-shell lines in addition to Fe L-shell
lines around 1 keV, and the \ion{Mg}{11} He-like 2$\rightarrow$1 line
complex at $\sim$1.34 keV.  The mere presence of these lines suggests
a moderately low electron temperature, $kT \sim 1$ keV, as our
detailed fits confirm.  Table 2 gives best-fit parameters and 1-sigma
statistical errors determined from two classes of NEI models: (1) a
``single ionization timescale'' model which assumes a single, constant
value for both the temperature and ionization timescale in the blast
wave region and (2) a ``planar shock'' model which also assumes a
constant temperature, but integrates the emission from spectral
components with ionization timescales varying linearly from 0 to
$n_et$ (the fit parameter) and weighted by equal fractional intervals
in $n_et$.  The first model provides a marginally acceptable fit to
the \chandra\ data (the $\chi^2$ of 57 for 38 degrees of freedom can
be rejected at the 97.5\% confidence level).  The second model, with
no additional parameters, yields a much improved, and statistically
acceptable, fit.

Elemental abundances were determined only for the species with obvious
line emission in the observed spectrum, viz.\ O, Ne, Mg, and Fe. The
abundances of the other astrophysically common species were fixed at
values appropriate to interstellar gas in the SMC (Russell \& Dopita
1992), an assumption that resulted in a significantly better fit than
one where the abundances of the other species were fixed to standard
solar abundances. The line-of-sight absorbing column density, $N_{\rm
H}$, was separated into a component with normal solar composition
arising from the Galaxy and a component from the SMC with lower
abundances. The Galactic component was fixed at an $N_{\rm H}$ value
consistent with reddening estimates and the Galactic \ion{H}{1} column
to \eohone, while the SMC component was allowed to vary freely.

The abundances from both model fits are nicely consistent with the
well-known low metallicity of the gas in the SMC ($\sim$25\% solar),
strongly suggesting that the emission we see from the outer rim comes
from the shock wave propagating through interstellar
gas.\footnotemark\footnotetext{One should be wary of interpreting the
fitted abundances much beyond this.  In particular, the Fe abundance
comes from a complex of L-shell lines that are unresolved at the
modest spectral resolution of our data.  Current models (including
ours) are known to be incomplete as regards the Fe L-shell emission
(Brickhouse et al.~2000), which introduces additional uncertainty
beyond the statistical errors in the measured value of the
Fe-abundance.}  Further in from the blast wave where the X-ray
emission of \eohone\ has become quite bright, the spectral character
has changed as well. In fig.~2 the spectrum from the bright region in
the southeast quadrant is plotted (labeled ``ejecta'').
The superposed model spectrum, shown for comparison, is the blast wave
NEI model scaled in intensity to fit the continuum level of the ejecta
spectrum.  There are clear differences between the ejecta and blast
wave spectra in terms of both thermodynamic state (note, for example,
the strength of the \ion{O}{7} He-like 2$\rightarrow$1 line complex at
$\sim$0.57 keV in the ejecta which is virtually absent in the blast
wave) and the elemental abundances (especially of O and Ne). The
enhanced abundances and less advanced ionization state of the ejecta
are consistent with expectations for emission from the reverse shock
propagating into higher density, metal-rich gas.

\section{DISCUSSION}

The X-ray expansion rate of \eohone\ is fully consistent with the
remnant's proper motion estimated from the location and radial
velocities of the oxygen-rich optical filaments. Specifically our
estimate of the remnant's age assuming free expansion,
$1000^{+340}_{-200}$ yr, agrees with that of Tuohy \& Dopita (1983).
At the outermost edge of the X-ray emission from \eohone\ ($R\approx
22^{\prime\prime} = 6.4 \rm \, pc$), the inferred velocity is
$6200^{+1500}_{-1600}$ km s$^{-1}$, which we consider in the following
to be the speed of the blast wave.  According to the Rankine-Huguniot
jump conditions in the absence of cosmic ray acceleration (Landau \&
Lifshitz 1976), this velocity corresponds to a mean post-shock
temperature of $kT_S = {3\over 16}\, \mu m_p v_S^2 = 45^{+25}_{-20}$
keV, using a mean mass per particle of $\mu =0.61$.  Our measurement
of the post-shock electron temperature from the \chandra\ X-ray
spectrum, $kT\ \lsim 1$ keV, is at least 25 times smaller than this
estimate.

There are two principle assumptions that go into our estimate for the
mean temperature: (1) that the post-shock thermal energy is
partitioned equally between electrons and ions, and (2) that only a
negligible fraction of the total energy of the shock goes into cosmic
rays.

It has been debated whether electron and ion temperatures are quickly
equilibrated by plasma processes behind collisionless high Mach number
SN shocks (McKee 1974; Cargill \& Papadopoulos 1988) or if the
electron and ion temperatures initially differ by their mass ratio
($m_p/m_e = 1836$) and then equilibrate slowly through Coulomb
collisions (Shklovskii 1968).  For \eohone\ we find, ignoring for the
moment any Coulombic heating of the electrons downstream, that the
ratio of electron thermal energy to upstream kinetic energy, $E_{\rm
e,th}/E_{\rm kin} = kT_e/({1\over 2}\,m_pv_S^2)\ \lsim1\%$, is much
smaller than the value of 12\% expected by Cargill \& Papadopoulos
(1988).

Downstream of the blast wave, electrons and ions (mainly protons) will
exchange energy through Coulomb interactions, which provide the
minimum level of heating expected in the case of nonequipartition.
The electron temperature varies as

$${dT_e \over dn_et} = 0.13 {T_p - T_e \over T_e^{3/2}}$$

\noindent
(in cgs units) (Spitzer 1978, p.~22), where we express the temperature
variation in terms of $n_et$, since we have an observational estimate
of this quantity from the X-ray spectral fits (Table 2).  This
equation predicts that the mean post-shock electron temperature
(averaging over $n_et$ from 0 to $4\times 10^{11}\rm\, cm^{-3}\, s$)
would be between 4.5 keV and 8 keV for proton temperatures
corresponding to the measured shock velocity (i.e., $T_p = 40$--120
keV). Note that the dependence of electron temperature on the actual
value of $n_et$ up to which the averaging is done is weak.  If the
lower limit on the measured $n_et$ value, $2\times 10^{11}\rm\,
cm^{-3}\, s$, from our spectral fits is used, then the average $kT_e$
range becomes 3.5--6 keV.  This estimate assumes that the total
pressure ($n_pT_p + n_eT_e$) remains constant in the post-shock
region, when, in fact, some adiabatic decompression should occur as
the shocked gas elements expand outward. Estimates of this effect,
assuming the post-shock gas acquires a speed of $3/4~v_S$, indicate
that the predicted electron temperatures may be reduced to no lower
than $\sim$2.5 keV, closer to, but still considerably higher than, the
measured blast wave electron temperature in \eohone.  These simple
models suggest that post-shock proton temperatures of 40--120 keV are
just too high and that they need to be considerably reduced in order
to match the \chandra\ measurements.  This in turn argues that cosmic
ray production is {\it not} negligible in \eohone\ and that the
relativistic particles may be absorbing a significant fraction of the
SN shock energy.

Recently Decourchelle et al.~(2000) have shown that the fraction of
energy going into cosmic rays is likely to be high in young SNRs and
that this effect has direct consequences for the thermal X-ray
emission. In such a situation, the shock jump conditions are modified
(e.g., Blandford \& Eichler 1987) so that for a given shock velocity
the compression ratio increases and the post-shock temperature
decreases. Nonlinear models of shock acceleration (Ellison 2000)
appear fully consistent with the observed shock parameters of \eohone.
In particular, these models predict a mean post-shock temperature of 1
keV (see figure 1 in Ellison 2000) for standard cosmic ray injection
efficiencies and a high Mach number shock, i.e., values of 100--300,
as appropriate for \eohone.  Thus it appears that efficient cosmic ray
acceleration alone is sufficient to explain our observations of
\eohone, even without the additional influence of nonequipartition of
electron and ion temperatures.

Unfortunately, a direct measurement of the post-shock proton
temperature in \eohone\ is not possible, due to the absence of any H
line emission from the remnant.  Therefore, constraints on the
efficiency of cosmic ray production in \eohone\ will need to come from
future detailed studies of the compression ratio, the structure of the
forward and reverse shocks, and the X-ray line spectrum using NEI
hydrodynamical models of SNRs that include cosmic-ray acceleration
(e.g., Decourchelle et al.~2000).  In addition, studies of other types
of remnants, in particular Balmer-dominated ones for which estimates
of the proton temperature are available from H$\alpha$ line profiles,
should prove helpful in further elucidating the physics of
collisionless shock fronts.

\acknowledgements

Helpful discussions with David Burrows, Don Ellison, Maxim
Markevitch, Jon Morse, and Wallace Tucker on the scientific content of
the article are gratefully acknowledged.  We appreciate Monique
Arnaud's support and hospitality during the course of this project.
This work was partially supported by NASA Grant NAG5-6420.

\vfill\eject

\newpage

\begin{deluxetable}{cccc}
\tablecaption{Observations of \eohone}
\tablewidth{4.5truein}
\tablehead{
\colhead{Observatory} & \colhead{Start Date} &
\colhead{Average MJD} & \colhead{Duration (s)} 
}
\startdata
{\it Einstein} (EHRI) & 1980 Apr 18 & 44347.5 & $22500$ \nl
{\it ROSAT}    (RHRI) & 1991 Nov 9  & 48598.2 & $21900$ \nl
{\it ACIS-S}          & 1999 Aug 23 & 51413.9 & $\phantom{0}8883$ \nl
\enddata
\end{deluxetable}

\newpage

\begin{deluxetable}{ccc}
\tablewidth{0pt}
\tablecaption{NEI Models of E0102.2-0.5\label{neimodels}}
\tablehead{
\colhead{} & \colhead{Single Ionization} & 
\colhead{}\\
\colhead{Parameters} & \colhead{Timescale} & 
\colhead{Planar Shock}
}
\startdata
$kT$ (keV)  
	& $0.48^{+0.08}_{-0.05}$
	& $0.78^{+0.16}_{-0.15}$ \\
$n_{e}t$ (cm$^{-3}$ s) 
	& $4^{+21}_{~-2}\times10^{11}$
        & $4^{+4}_{-2}\times10^{11}$\\
$N_{\mathrm{H,SMC}}$\tablenotemark{a}~  (cm$^{-2}$)  
        & $(5\pm 3)\times10^{20}$
	& $<2.5\times10^{20}$ \\
%$N_{\mathrm{H,Gal}}$\tablenotemark{b}~  (cm$^{-2}$)
%        & $5\times10^{20}$ & $5\times10^{20}$ \\
% Emission Integral & & \\
$n_{\mathrm{e}}n_{\mathrm{H}}$V/4$\pi$D$^{2}$ (cm$^{-5}$)
	& $1.1^{+0.4}_{-0.3}\times10^{11}$
	& $4.1^{+2.0}_{-0.8}\times10^{10}$ \\
Oxygen 
	& $0.26^{+0.19}_{-0.14}$
	& $0.34^{+0.15}_{-0.12}$ \\
Neon
	& $0.31^{+0.11}_{-0.09}$
	& $0.67^{+0.20}_{-0.19}$\\
Magnesium
	& $0.28^{+0.12}_{-0.10}$
	& $0.47\pm0.16$ \\
Iron 	
	& $0.01\pm0.005$
	& $0.06\pm0.03$ \\
$\chi ^{2}$ (d.o.f.) 
        & 57.13 (38) & 46.07 (38) \\
\enddata
\tablenotetext{a}{Uses mean SMC abundances; fits also include a fixed 
$N_{\mathrm{H,Gal}}$ of $5\times10^{20}$ cm$^{-2}$ using solar abundances}
%\tablenotetext{b}{Fixed, uses solar abundances}
\end{deluxetable}

\newpage

\clearpage

\figcaption[f1.ps]{Global mean percentage expansion of SNR \eohone\
versus the time difference between imaging measurements. The point
indicating the shortest time difference comes from comparing the {\it
ROSAT} and \chandra\ observations. The other point comes from
comparing the {\it Einstein} and \chandra\ observations.  The error
bars show the statistical uncertainty, while the boxes that surround
each data point give an estimate of the systematic uncertainty.  The
dotted curves show the best-fit percentage expansion and 1 sigma
errors.  The inferred speed of the shock front is $\sim$6000 km
s$^{-1}$ and the remnant's age is 1000 yr.
\label{Figure 1}}

\figcaption[f2.ps]{\chandra\ spectra of a portion of the outer blast
wave and bright ejecta of SNR \eohone\ as indicated.  The spectra were
rebinned to a minimum of 25 events per channel. The best-fit
nonequilibrium ionization model for the blast wave is shown as the
solid histogram.  The model appears compared against the ejecta
spectrum as well in order to demonstrate the gross spectral
differences between the two regions. The residuals are shown only for
the blast wave spectrum.
\label{Figure 2}}

%\end{document}

\clearpage

\begin{figure}
\plotfiddle{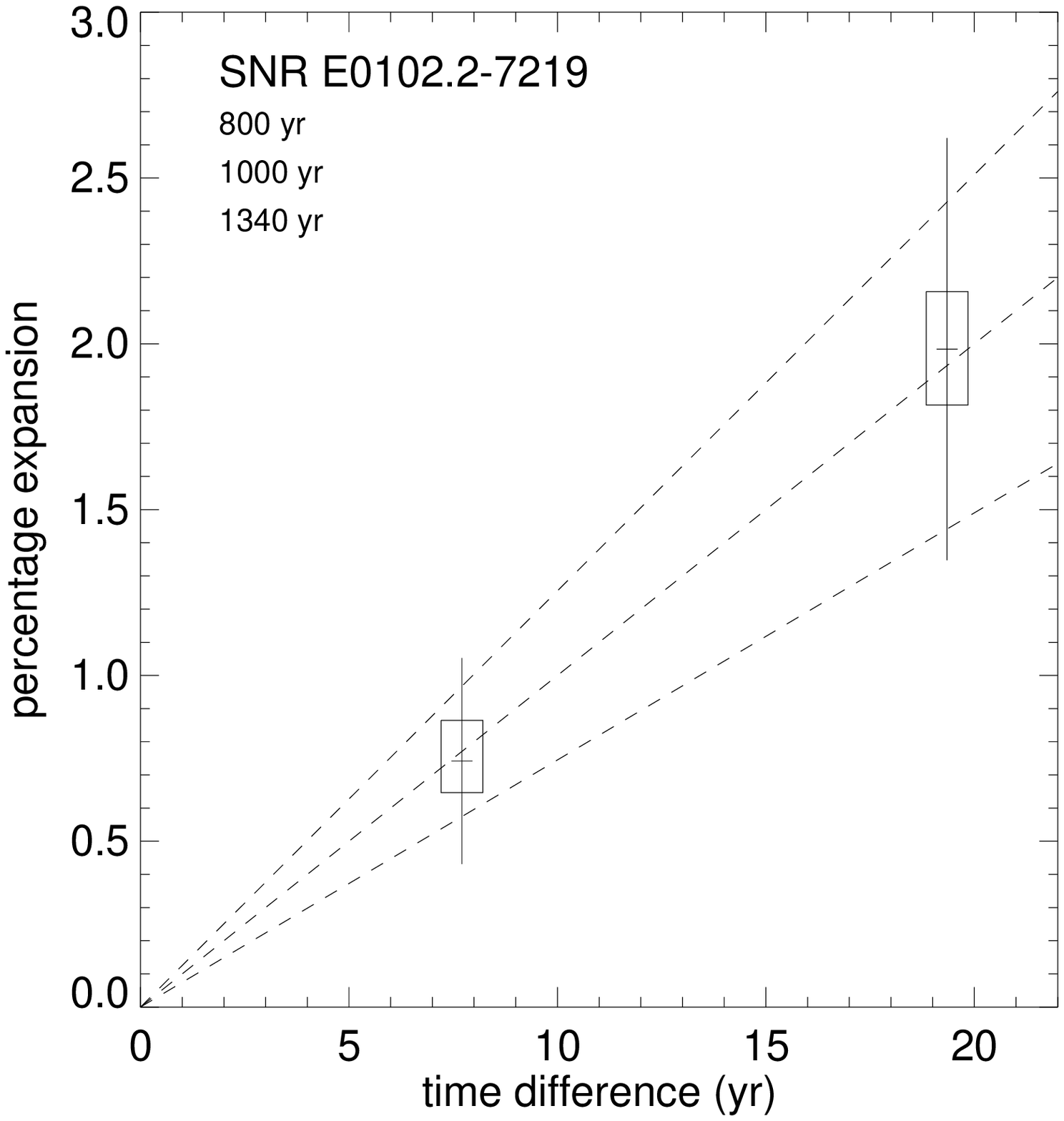}{5in}{0}{100}{100}{-300}{-200}
\end{figure}

\clearpage

\begin{figure}
\plotfiddle{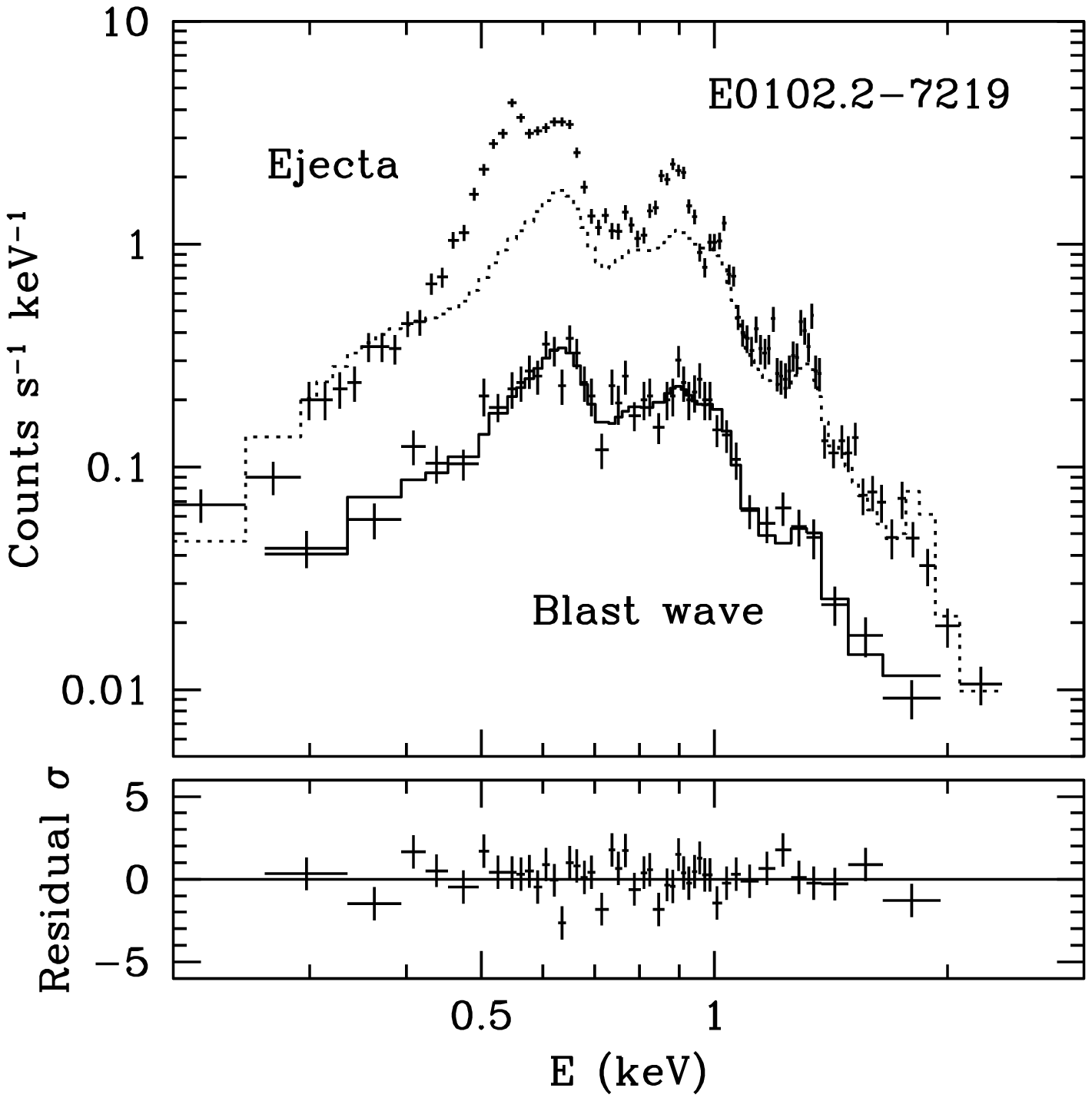}{5in}{0}{100}{100}{-300}{-200}
\end{figure}

\end{document}